\documentclass[fleqn,usenatbib]{mnras}

\usepackage{newtxtext,newtxmath}

\usepackage[T1]{fontenc}
\usepackage{ae,aecompl}
\usepackage{lscape}

\usepackage{graphicx}	
\usepackage{amsmath}	
\usepackage{amssymb}	

\newcommand{\tess}{{\it TESS}}

\newcommand{\NGTS}{NGTS}
\newcommand{\TESS}{{\it TESS}}

\newcommand{\kms}{km\,s$^{-1}$}

\newcommand{\masy}{mas\,y$^{-1}$}

\newcommand{\msun}{\mbox{$M_{\odot}$}}
\newcommand{\rsun}{\mbox{$R_{\odot}$}}

\newcommand{\Nstar}{TOI-222}
\newcommand{\Ngamma}{\mbox{$13.51423 \pm 0.005$}}
\newcommand{\NRA}{\mbox{$23^{\rmn{h}} 56^{\rmn{m}} 38\fs99$}} 
\newcommand{\NDec}{\mbox{$-44\degr 43\arcmin 09\farcs 81$}} 
\newcommand{\NpropRA}{\mbox{$161.769\pm0.049 $}} 
\newcommand{\NpropDec}{\mbox{$-75.541.0\pm0.055$}} 

\newcommand{\Nteff}{\mbox{$5815 \pm 85$}}
\newcommand{\Nmetal}{\mbox{$0.11 \pm 0.06$}} 
\newcommand{\Ndist}{\mbox{$84\,^{+0.3}_{-0.3}$}}
\newcommand{\Nlogg}{\mbox{$4.5 \pm 0.13$}}

\newcommand{\Nvsini}{$3.60 \pm 1.35$}
\newcommand{\NVmag}{$ 9.528 \pm 0.37$}
\newcommand{\NBmag}{$10.009 \pm 0.013$}

\newcommand{\Ngmag}{$9.609 \pm 0.03$}
\newcommand{\Nrmag}{$9.424 \pm 0.392$}
\newcommand{\Nimag}{$9.333 \pm 0.412$}
\newcommand{\NGAIAmag}{$9.15167\pm 0.00035$}

\newcommand{\NTmag}{$8.723 \pm 0.017$}
\newcommand{\NHmag}{$7.869 \pm  0.024$}
\newcommand{\NJmag}{$8.144\pm 0.019$}
\newcommand{\NKmag}{$7.7729 \pm 0.021$}
\newcommand{\NWmag}{$7.681 \pm  0.029$}
\newcommand{\NWWmag}{$7.763 \pm 0.019$}

\newcommand{\Nplanet}{TOI-222}

\newcommand{\Nperiodshort}{\mbox{$33.9$}}

\newcommand\T{\rule{0pt}{2.2ex}}

\title[\Nstar]{\Nstar:  a single-transit \tess\ candidate revealed to be a 34-day eclipsing binary with CORALIE, EulerCam and NGTS}

\author[M. Lendl et al.]{
\parbox{\textwidth}{
Monika Lendl,$^{1,2}$\thanks{E-mail: \href{monika.lendl@unige.ch}{monika.lendl@unige.ch}}
Fran\c{c}ois Bouchy,$^{1}$
Samuel Gill,$^{3,4}$
Louise~D.~Nielsen,$^{1}$
Oliver Turner,$^{1}$
Keivan Stassun,$^{5,6}$
Jack S. Acton,$^{7}$
David R. Anderson,$^{3,4}$
David J. Armstrong,$^{3,4}$
Daniel~Bayliss,$^{3,4}$
Claudia Belardi,$^{7}$
Edward M. Bryant,$^{3,4}$
Matthew R. Burleigh,$^{7}$
Alexander Chaushev,$^{8}$
Sarah L. Casewell,$^{7}$
Benjamin F. Cooke,$^{3,4}$
Philipp~Eigm\"uller,$^{8}$
Edward~Gillen,$^{9}$
Michael R.~Goad,$^{7}$
Maximilian~N.~G{\"u}nther,$^{10,11}$
Janis Hagelberg,$^{1}$
James~S.~Jenkins,$^{12,13}$
Tom Louden,$^{3,4}$
Maxime Marmier,$^{1}$
James~McCormac,$^{3,4}$
Maximiliano Moyano,$^{14}$
Don~Pollacco,$^{3,4}$
Liam Raynard,$^{7}$
Rosanna H. Tilbrook,$^{7}$
St\'{e}phane~Udry,$^{1}$
Jose I. Vines,$^{12}$
Richard~G.~West,$^{3,4}$
Peter~J.~Wheatley,$^{3,4}$
George Ricker,$^{10}$ 
Roland Vanderspek,$^{10}$ 
David W. Latham,$^{15}$ 
Sara Seager,$^{10,16,17}$  
Josh Winn,$^{18}$ 
Jon M. Jenkins,$^{19}$ 
Brett Addison,$^{20}$ 
C\'{e}sar Brice\~{n}o,$^{21}$ 
Rafael Brahm,$^{22,23}$   
Douglas A. Caldwell,$^{19,24}$ 
John Doty,$^{25}$ 
N\'estor Espinoza,$^{26,27}$ 
Bob Goeke,$^{10}$ 
Thomas Henning,$^{27}$ 
Andr\'es Jord\'an,$^{23,28}$ 
Akshata Krishnamurthy,$^{10}$ 
Nicholas Law,$^{29}$ 
Robert Morris,$^{19,24}$ 
Jack Okumura,$^{20}$ 
Andrew W. Mann,$^{29}$ 
Joseph E. Rodriguez,$^{15}$ 
Paula Sarkis$^{27}$ 
Joshua Schlieder,$^{30}$ 
Joseph D. Twicken,$^{19,24}$ 
Steven Villanueva Jr.,$^{10,31}$ 
Robert A. Wittenmyer,$^{20}$ 
Duncan J. Wright,$^{20}$ 
Carl Ziegler,$^{32}$ 
}
\\
$^{1}$Observatoire de Gen{\`e}ve, Universit{\'e} de Gen{\`e}ve, 51 Ch. des Maillettes, 1290 Sauverny, Switzerland\\ 
$^{2}$Space Research Institute, Austrian Academy of Sciences, Schmiedlstr. 6, 8042 Graz, Austria\\
$^{3}$Dept.\ of Physics, University of Warwick, Gibbet Hill Road, Coventry CV4 7AL, UK\\
$^{4}$Centre for Exoplanets and Habitability, University of Warwick, Gibbet Hill Road, Coventry CV4 7AL, UK\\
$^{5}$Vanderbilt University, Deparment of Physics \& Astronomy, 6301 Stevenson Center Ln., Nashville, TN 37235, USA\\
$^{6}$Fisk University, Department of Physics, 1000 18th Ave. N., Nashville, TN 37208, USA\\
$^{7}$Department of Physics and Astronomy, University of Leicester, University Road, Leicester, LE1 7RH, UK\\
$^{8}$Institute of Planetary Research, German Aerospace Center, Rutherfordstrasse 2, 12489 Berlin, Germany\\
$^{9}$Astrophysics Group, Cavendish Laboratory, J.J. Thomson Avenue, Cambridge CB3 0HE, UK\\
$^{10}$Department of Physics, and Kavli Institute for Astrophysics and Space Research, Massachusetts Institute of Technology,Cambridge, MA 02139, USA\\
$^{11}$Juan Carlos Torres Fellow\\
$^{12}$Departamento de Astronomia, Universidad de Chile, Casilla 36-D, Santiago, Chile\\
$^{13}$Centro de Astrof\'isica y Tecnolog\'ias Afines (CATA), Casilla 36-D, Santiago, Chile.\\
$^{14}$Instituto de Astronom\'ia, Universidad Cat\'{o}lica del Norte, Angamos 0610, 1270709, Antofagasta, Chile\\
$^{15}$Center for Astrophysics, Harvard \& Smithsonian, 60 Garden St, Cambridge, MA 02138, USA\\
$^{16}$Department of Earth, Atmospheric and Planetary Sciences, Massachusetts Institute of Technology, Cambridge, MA 02139, USA\\
$^{17}$Department of Aeronautics and Astronautics, MIT, 77 Massachusetts Avenue, Cambridge, MA 02139, USA\\
$^{18}$Department of Astrophysical Sciences, Princeton University, Princeton, NJ 08544, USA\\
$^{19}$NASA Ames Research Center, Moffett Field, CA 94035, USA\\
$^{20}$University of Southern Queensland, Centre for Astrophysics, West Street, Toowoomba, QLD 4350 Australia\\
$^{21}$Cerro Tololo Inter-American Observatory, Casilla 603, La Serena, Chile\\
$^{22}$Center of Astro-Engineering UC, Pontificia Universidad Cat\'olica de Chile, Av. Vicu\~{n}a Mackenna 4860, 7820436 Macul, Santiago, Chile\\
$^{23}$Millennium Institute for Astrophysics, Chile\\
$^{24}$SETI Institute, 189 Bernardo Ave., Suite 200, Mountain View, CA 94043, USA\\
$^{25}$Noqsi Aerospace Ltd., 15 Blanchard Avenue, Billerica, MA, 01821, USA\\
$^{26}$Space Telescope Science Institute, Baltimore, USA\\
$^{27}$Max-Planck-Institut f\"ur Astronomie, K\"onigstuhl 17, Heidelberg 69117, Germany\\
$^{28}$Facultad de Ingenier\'ia y Ciencias, Universidad Adolfo Ib\'a\~nez, Av.\ Diagonal las Torres 2640, Pe\~nalol\'en, Santiago, Chile\\
$^{29}$Department of Physics and Astronomy, The University of North Carolina at Chapel Hill, Chapel Hill, NC 27599-3255, USA\\
$^{30}$NASA Goddard Space Flight Center, 8800 Greenbelt Road, MD, USA\\
$^{31}$Pappalardo Fellow
$^{32}$Dunlap Institute for Astronomy and Astrophysics, University of Toronto, 50 St. George Street, Toronto, Ontario M5S 3H4, Canada\\
}
\date{in prep.}

\pubyear{20NN}

\begin{document}
\label{firstpage}
\pagerange{\pageref{firstpage}--\pageref{lastpage}}
\maketitle

\begin{abstract}
We report the period, eccentricity, and mass determination for the \tess\ single-transit event candidate \Nstar, which displayed a single 3000\,ppm transit in the \tess\ two-minute cadence data from Sector 2.  We determine the orbital period via radial velocity measurements (P=\Nperiodshort\,days), which allowed for ground-based photometric detection of two subsequent transits.  Our data show that the companion to \Nstar\ is a low mass star, with a radius of $0.18_{-0.10}^{+0.39}$\,\rsun and a mass of $0.23\pm0.01$\,\msun. This discovery showcases the ability to efficiently discover long-period systems from \tess\ single transit events using a combination of radial velocity monitoring coupled with high precision ground-based photometry.  
\end{abstract}




\section{Introduction}  \label{sec:intro}
Even though the probability of an exoplanet to transit across its host plummets for long orbital periods, space-based transit surveys such as CoRoT \citep{Baglin06} and Kepler \citep{Borucki10} have succeeded in detecting a number of long-period transiting planets as they monitor large fields continuously for extended periods of time. Unlike its predecessors, the Transiting Exoplanet Survey Satellite (\TESS, \citealt{ricker2014}) is optimised for bright stars. From its recently-completed survey of the Southern ecliptic hemisphere, over 1000 promising exoplanet candidates, including multi-planet candidate systems, have been reported as \TESS\ objects of interest (TOIs).  A number of the candidates have now also been confirmed as bona fide transiting planets via radial velocity measurements; for example pi~Mensae\,c \citep{huang19,gandolfi18}, HD~202772A\,b \citep{wang19}, and HD~1397\,b \citep{nielsen19,brahm19}.

As predicted by simulations \citep{cooke18,villanueva19}, TESS also detected a number of promising exoplanet candidates for which only a single transit event was observed. These "Monotransits" do not provide well-determined periods, although period estimates can be derived using knowledge of the host star and transit parameters \citep{osborn16}. Since the time coverage of \TESS \, is limited to 27~days for large parts of the sky, these single-transit event candidates provide longer period planetary systems and eclipsing binaries. The confirmation of monotransit candidates through the observation of a second transit is challenging as the planetary period is only loosely constrained. This task can be made easier by monitoring the primary's radial velocity in order to constrain the spectroscopic orbit and predict the times of subsequent transit events.

In this work, we present the mass and period determination of \Nplanet: a TOI identified as a \TESS \, single-transit event in Sector 2. Using spectroscopic follow-up, we determined the orbit of this low-mass eclipsing binary, leading to the recovery of the transit from subsequent photometric observations. We describe the observations and the analysis of our data in Sections \ref{sec:obs} and \ref{sec:ana}, and place this detection into context in Section \ref{sec:conc}.


\section{Observations} \label{sec:obs}
\subsection{\tess\ Discovery Photometry} \label{sec:tess}

The V=9.3 star HD~224286 (TIC~144440290) was observed by \tess\ in Sector~2 throughout 27.4~days. The observations have a continuous 2-minute cadence, with only a short gap occurring at the centre of the sequence when observations were interrupted to allow for the data to be downlinked to Earth. A single, approximately 3.5~mmag deep, transit event was identified and denominated \Nplanet\ by the \tess\ team. The transit occurred approximately 22~days after the start of the sequence, thus constraining the period of a potential planet or binary companion to be above 22~days. The long period but comparatively short total transit duration of 1.8\,hours, paired with a clear V shape, indicated a grazing configuration, with a larger planet or an eclipsing binary only partially transiting the star.

\begin{figure*}
	\includegraphics[width=\linewidth]{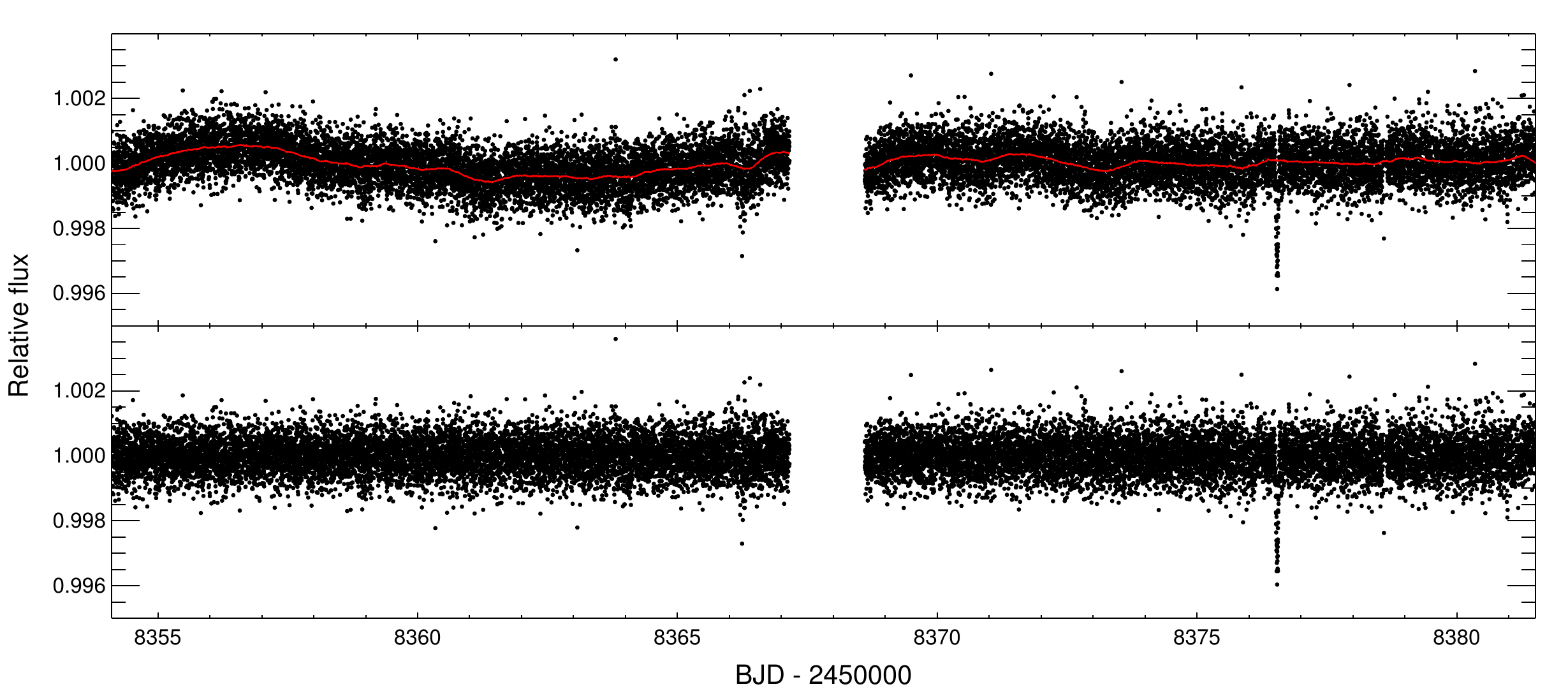}
    \caption{TESS data of TOI-222. The top panel shows the uncorrected data in black with the 10-hour boxcar applied to correct for stellar and instrumental systematics in red. The bottom panel shows the corrected TESS data.}
    \label{fig:tessphot}
\end{figure*}

We used the TESS Science Processing Operations Center \citep{Jenkins2016} two-minute cadence PDC light curve \citep{Smith2012,Stumpe2014} for the subsequent analysis, and additionally filtered the time-series by a 12-hour boxcar to remove a low-amplitude variation likely caused by stellar active regions rotating in and out of view. Data points obtained in transit were excluded at this step in order to avoid altering the transit shape when correcting these long-term trends. The \tess\ data are shown in Figure~\ref{fig:tessphot}. 

\subsection{High resolution spectroscopy with CORALIE}
\label{sub:spect}
\begin{figure}
	\includegraphics[width=\columnwidth]{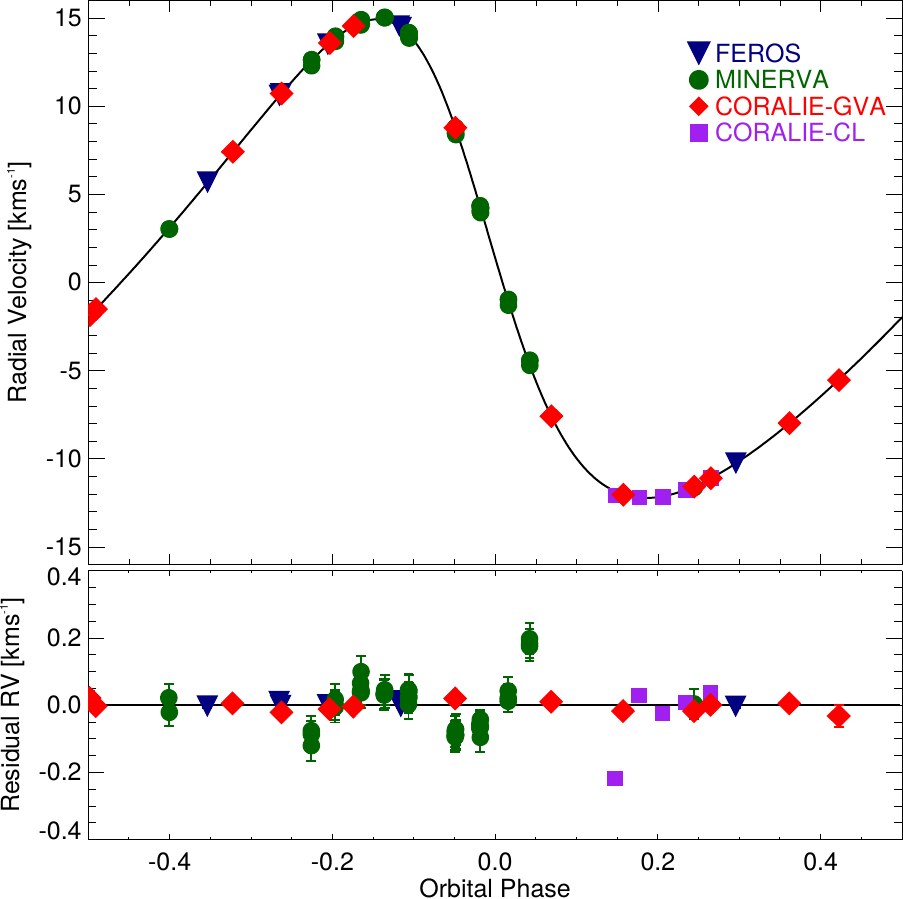}
    \caption{\label{fig:RV}Radial velocities of TOI-222 together with the best-fit RV solution (top panel), and best-fit residuals (bottom panel). CORALIE data from the Geneva (``CORALIE-GVA'') and the MPIA-Chile (``CORALIE-CL'') teams were reduced with different pipelines and are shown as red diamonds and purple squares, respectively; FEROS data are shown as blue triangles, and MINERVA data are shown in green. Error bars are smaller than the size of the data points in most cases.}
\end{figure}

TOI-222 was observed with the CORALIE spectrograph on the Swiss 1.2 m Euler telescope at La Silla Observatory, Chile \citep{CORALIE} between 2 January and 30 June 2019. 
CORALIE has a resolving power of $R\sim60000$ and is fed by two fibres; one 2\arcsec\ on-sky science fibre encompassing the star and another which can either 
be connected to a Fabry-P\'{e}rot etalon for simultaneous wavelength calibration or on-sky for background subtraction of the sky-flux. Radial-velocities (RVs) were computed for each epoch by cross-correlating with a binary G2 mask \citep{Pepe2002}. Bisector-span, FWHM and other line-profile diagnostics were computed 
as well using the standard CORALIE pipeline. As initial observations showed a large 10 km/s RV shift over 9 days, we reduced the exposure time from 1800 seconds to 900-300 seconds depending on seeing and airmass to save telescope time. We obtain a final precision of 6-20 m/s. The resulting velocities are given in Table~\ref{tab:rvs} and are shown in Figure \ref{fig:RV}.
The CORALIE spectra were shifted to the stellar rest frame and stacked while weighting the contribution from each spectrum with its mean flux to produce a high signal-to-noise spectrum for spectral characterisation.

We detected significant RV variations in the CORALIE data, indicative of a low-mass stellar component orbiting the main target with a period of 33.9 days, which prompted us to proceed with photometric follow-up.

\begin{table*}
	\centering
	\caption{Radial Velocities for \Nstar.}
	\label{tab:rvs}
	\begin{tabular}{rrrrrr} 
BJD			&	RV		&RV err &	FWHM&BIS&Instrument\\
(-2400000)	& (\kms)& (\kms)&(\kms) &(\kms) \\
		\hline
58486.571350  & 1.94699 &	0.01589 &	8.19162 &	-0.03997 & CORALIE\\
58490.540095 &	5.54885 &	0.00645 &	8.25366 &	-0.05706 & CORALIE\\
58495.549674 &	12.01076 &	0.00647 &	8.26060 &	-0.05091 & CORALIE\\
58510.528803 &	22.29837 &	0.01479 &	8.23169 &	-0.00504 & CORALIE\\
58514.527724 &	5.94552 &	0.01239 &	8.26212 &	-0.04074 & CORALIE\\
58517.523825 &	1.48704 &	0.01297 &	8.27996 &	-0.05966 & CORALIE\\
58526.513423 &	7.98638 &	0.03271 &	8.31453 &	-0.09361 & CORALIE\\
58636.901387 &	20.93734 &	0.01824 &	8.24319 &	-0.09194 & CORALIE\\
58638.938599 &	24.24232 &	0.01393 &	8.23949 &	-0.00828 & CORALIE\\
58640.945601 &	27.10596 &	0.01844 &	8.27672 &    0.00445 & CORALIE\\
58641.935281 &	28.06841 &	0.01544 &	8.24595 &	-0.06231 & CORALIE\\
58656.822502 &	2.42260 &	0.01350 &	8.31897 &	-0.02812 & CORALIE\\
58664.859362 &	11.6754 &	0.02199 &	8.23540 &	-0.00685 & CORALIE\\
		\hline
	\end{tabular}
\end{table*}

\subsection{Transit Recovery With EulerCam}
\label{sub:eulerphot}

With the system period established, we scheduled photometric observations with EulerCam, also installed at the 1.2~m Euler telescope, with the goal of detecting a second transit of the low-mass companion in front of its host. The observations were carried out throughout 2.8~hours on 13 June 2019, covering approximately 75\% of the predicted 1-$\sigma$ transit window. We used an Cousins I filter, 20\,s exposures and applied a large defocus to avoid saturating the target while keeping a reasonable observation efficiency.  For details on the instrument and the data analysis routines used to extract relative aperture photometry, please refer to \citet{Lendl12}. We detected a flux drop of $\sim 3.5$~mmag during first half of the sequence, having a depth compatible with that of the TESS measurement. The egress occurred approx. 42 minutes (or $3.8 \sigma$) earlier than predicted, indicating a slightly shorter planetary period. The EulerCam light curve is shown in Figure~\ref{fig:LCs}, and the dataset is presented in Table \ref{tab:ecam}.

\subsection{Transit Confirmation with \NGTS{}}
\label{sub:ngtsphot}

TOI-222 was initially scheduled for a blind search with \NGTS\ to recover the orbital period following its designations as a single-transit object of interest from \TESS. We obtained 37 nights of single-telescope observations (18,510 images with 10\,s exposures). 

Once the system's period was known, we scheduled NGTS observations to cover the next primary eclipse of TOI-222, occurring on 17 Jul 2019. In total, 10 telescopes were used to simultaneously observe the primary eclipse, collecting over 10,000 individual data points with an exposure time of 10\,seconds each. The data were reduced using the CASUTools\footnote{\url{http://casu.ast.cam.ac.uk/surveys-projects/software-release}} photometry package and we applied the SysRem algorithm \citep{Tamuz2005} to detrend the data. For more details on NGTS, please refer to \citet{ngts}. The resulting light curve was binned to 2 minutes and detrended using a third-order time polynomial, obtaining an RMS of $\sim$450\,ppm .

\begin{figure}
	\includegraphics[width=\columnwidth]{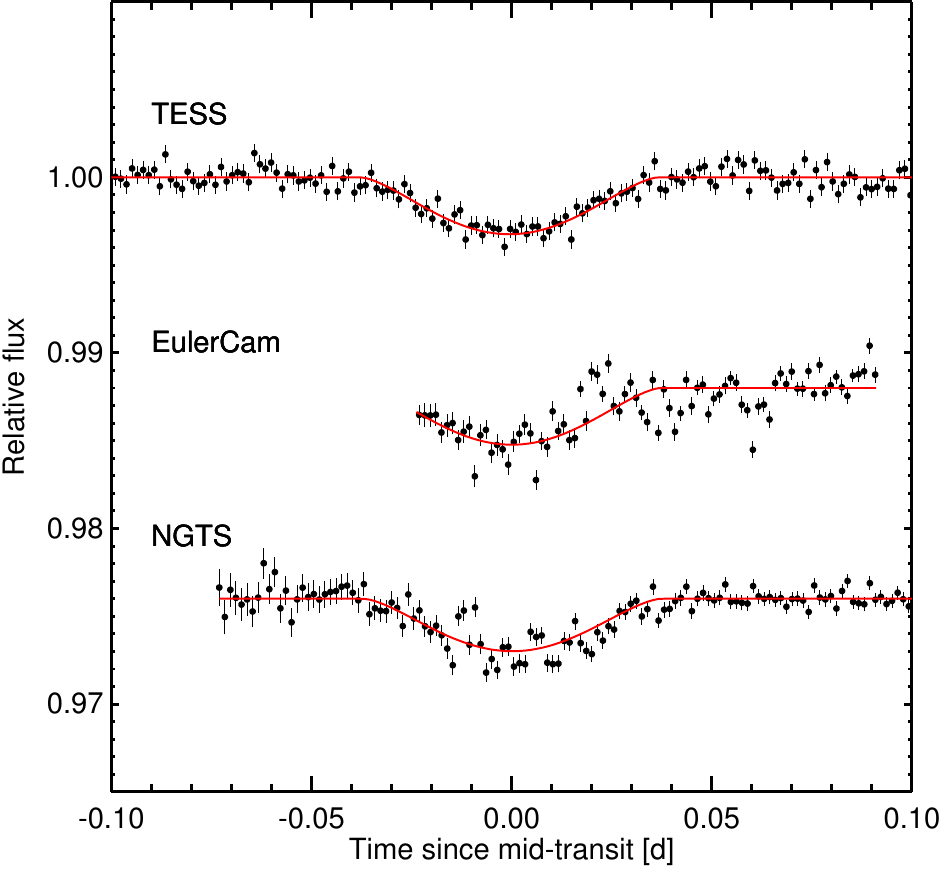}
    \caption{Photometric data used to perform a global fit of the system. These are (from top to bottom) \TESS, EulerCam and NGTS. All light curves have been detrended and binned into 2-minute intervals.}
    \label{fig:LCs}
\end{figure}

We clearly detect the primary eclipse of TOI-222 with NGTS, confirming the slightly shorter period and distinct V-shape of the eclipse. The NGTS data are shown in Figure \ref{fig:LCs} and the data are given in Table \ref{tab:ngts}.

\begin{table*}
	\centering
	\caption{\label{tab:ecam}EulerCam photometry. The full table is available from the online version of this paper.}
	\begin{tabular}{rrrrrrrrr} 
BJD		&	Flux	& Flux error & xshift & yshift & airamass & FWHM & sky flux & exposure time\\
(-245000) &  & & pixel & pixel & & pixel & $e^{-}$ & s \\
\hline
8647.819731 & 0.996430  & 0.001189 & -0.76 &  2.31 & 1.41 & 14.09 &   42.96  &  20.00 \\
8647.820159 & 0.993507  & 0.001186 & -0.45 &  3.02 & 1.41 & 15.29 &   42.88  &  20.00\\
8647.820589 & 0.998077  & 0.001191 & -1.33 &  1.81 & 1.41 & 14.37 &   42.58  &  20.00\\
8647.821005 & 0.995025  & 0.001175 &  0.24 &  0.92 & 1.40 & 15.59 &   42.72  &  20.00\\
8647.821431 & 0.994832  & 0.001174 &  3.46 & -1.46 & 1.40 & 15.77 &   41.77  &  20.00\\
		\hline
	\end{tabular}
\end{table*}

\begin{table}
	\centering
	\caption{\label{tab:ngts}NGTS photometry, binned into 2 minute intervals. The full table is available from the online version of this paper.}
	\begin{tabular}{rrr} 
BJD			&	Flux	& Flux error\\
\hline
2458681.682801  & 1.000619 & 0.001027  \\
2458681.684190  & 0.998942 & 0.000992 \\
2458681.685579  & 1.000491 & 0.000973 \\
2458681.686967  & 1.000034 & 0.000956 \\
2458681.688356  & 0.999649 & 0.000928 \\
		\hline
	\end{tabular}
\end{table}

\subsection{Additional high-resolution spectroscopy}

\subsubsection{High resolution spectroscopy with FEROS and CORALIE}
The radial velocity variation of TOI-222 was also monitored
using the FEROS spectrograph \citep[R=48000][]{kaufer:99} mounted on the MPG 2.2m telescope installed in La Silla Observatory. A total number
of seven spectra were obtained between June 4 and July 13 of 2019. The adopted exposure time was of 300s which produced spectra with a typical SNR per resolution element of 130. Observations were performed using the simultaneous calibration mode in which a comparison fiber is illuminated with a Thorium-Argon lamp in order to trace the instrumental variations during the science exposure. Next to these data, we also obtained an additional four data points with CORALIE under Chilean time. These data were processed with the automated CERES package \citep{ceres} which performs the optimal extraction, wavelength calibration, and computation of precision radial velocities using the cross-correlation technique.

\subsubsection{High resolution spectroscopy with MINERVA-Australis}
A total of 60 spectra of TOI-222 were obtained at 15 epochs between 2019 May 11 and July 18, using the MINERVA-Australis telescope array at Mt. Kent Observatory in Queensland, Australia \citep{witt18,addison19}.  All of the telescopes in the MINERVA-Australis array simultaneously feed a single Kiwispec R4-100 high-resolution (R~80,000) spectrograph with wavelength coverage from 500 to 630 nm over 26 echelle orders.  Light from each of the four telescopes is delivered to individual fibres (numbered 3, 4, 5, and 6), and calibration is achieved via a simultaneous Thorium-Argon lamp illuminating fibres 1 and 7.  We derived radial velocities for each telescope using the least-squares analysis of \citet{ang12} and corrected for spectrograph drifts with simultaneous Thorium Argon arc lamp observations.  TOI-222 was observed simultaneously with up to four telescopes, with one or two 30-minute exposures per epoch.  The radial velocities from each telescope are given in Table \ref{tab:rvs} labeled by their fibre number.  Each telescope (fibre) has its own velocity zero-point which is modelled as a free parameter, and the mean uncertainty of the 60 MINERVA-Australis observations is 4.5\,\mbox{m\,s$^{-1}$}. Additional scatter in the MINERVA data was introduced by a power cut at the observatory.

\subsection{Speckle imaging}

\begin{figure}
	\includegraphics[width=\columnwidth]{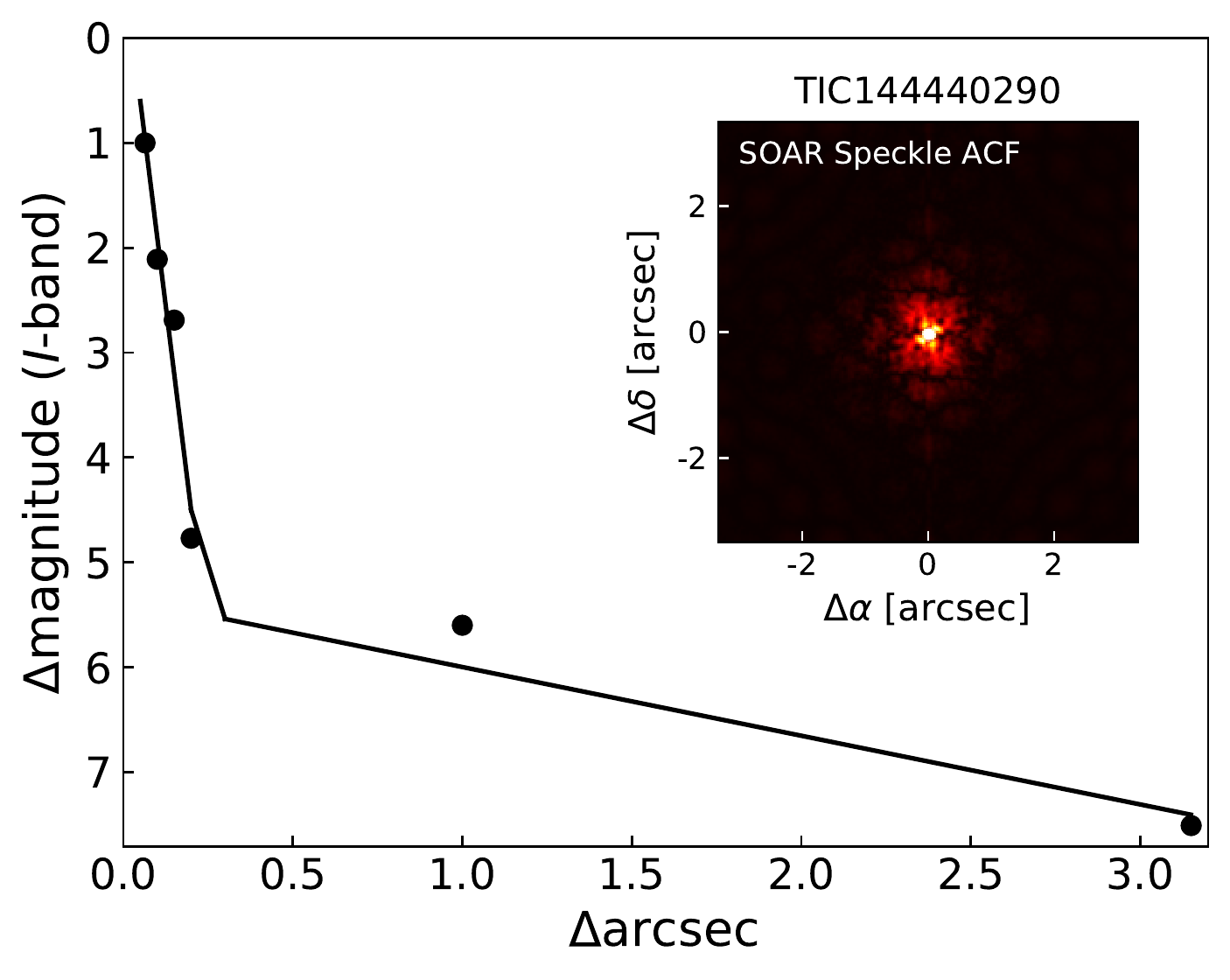}
    \caption{The 5$\sigma$ detection sensitivity to companion stars to TOI-222 for the SOAR speckle observation. Inset is the speckle auto-correlation function derived from the SOAR observation. No stars were detected within 3\arcsec of TOI-222.}
    \label{fig:speckle}
\end{figure}

Nearby stars which fall within the same 21\arcsec\ TESS pixel as TOI-222 could contaminate the \TESS\ photometry, resulting in a reduced transit depth. We searched for unaccounted companions to TOI-222 with SOAR speckle imaging \citep{tokovinin18} on 18 May 2019 UT, observing in a similar visible bandpass as \TESS. Additional details of the observation are available in \citet{ziegler19}. We detected no nearby stars within 3\arcsec of TOI-222. The 5$\sigma$ detection sensitivity and the speckle auto-correlation function from the SOAR observation are plotted in Figure \ref{fig:speckle}.

\section{System Analysis}
\label{sec:ana}

\subsection{Properties of the Primary}
\label{sec:stellar}

We used the CORALIE spectra to derive atmospheric parameters for the primary. Each spectrum was corrected into the laboratory reference frame and co-added onto a common wavelength range. Maximum and median filters were applied to identify continuum regions which were fitted with spline functions (one every nm) to normalise the spectra \citep[a standard function within ispec v20161118; ][]{2014A&A...569A.111B}. We used wavelet analysis \citep{2018A&A...612A.111G} to determine the best-fitting atmospheric parameters for the host star, which works by comparing the wavelet coefficients of the target spectrum to a grid of model spectra.  We re-sample between 450-650\,nm with 2$^{17}$ values and co-add the spectra. We calculate the wavelet coefficients Wi=4-14 and fit the same coefficients with model spectra (identical to those used in \citealt{2018A&A...612A.111G} and \citealt{2019A&A...626A.119G}) in a Bayesian framework. We initiated 100 walkers and generated 10,000 draws as a burn-in phase. We generated a further 10,000 draws to sample the posterior probability distribution (PPD) for the stellar effective temperature ($T_{\rm eff}$), metallicity ([Fe/H]), projected rotational velocity ($V\sin\,i$) and surface gravity ($\log g$). \citet{2018A&A...612A.111G} note an [Fe/H] offset of -0.18\,dex, which we correct for by adding 0.18\,dex to the PPD for [Fe/H]. The wavelet method for CORALIE spectra can determine $T_{\rm eff}$ to a precision of 85\,K, [Fe/H] to a precision of 0.06\,dex and $V\sin\,i$ to a precision of 1.35\,km\,s$^{-1}$ for stars with $V\sin\,i$ $\geq$ 5\,km\,s$^{-1}$. However, measurements of log g from wavelet analysis are not reliable beyond confirming dwarf-like gravity (log g $\approx$ 4.5\,dex). Subsequently, we fit the wings of the magnesium triplets with spectral synthesis by fixing $T_{\rm eff}$, [Fe/H] and $V\,\sin\,i$ and changing log g until an acceptable fit was found. The inferred values are listed in Table \ref{tab:stellar}.

For an empirical determination of the stellar radius, we performed an analysis of the broadband spectral energy distribution (SED) together with the {\it Gaia\/} parallax, following the procedures described in \citet{Stassun:2016,Stassun:2017,Stassun:2018}. We pulled the $B_T V_T$ magnitudes from {\it Tycho-2}, the $BVgri$ magnitudes from APASS, the $JHK_S$ magnitudes from {\it 2MASS}, the W1--W4 magnitudes from {\it WISE}, the $G$ magnitude from {\it Gaia\/}, and the FUV and NUV fluxes from {\it GALEX}. Together, the available photometry spans the full stellar SED over the wavelength range 0.15--22~$\mu$m. 
We performed a fit using Kurucz stellar atmosphere models, with the priors on $T_{\rm eff}$, $\log g$, and [Fe/H] from the spectroscopic analysis. The remaining free parameter is the extinction ($A_V$), which we limited to the maximum line-of-sight extinction from the \citet{Schlegel:1998} dust maps. The resulting fit is good with a reduced $\chi^2$ of 3.9, and a best fit extinction of $A_V = 0.02 \pm 0.02$. Integrating the (unextincted) model SED gives the bolometric flux at Earth of $F_{\rm bol} = 5.15 \pm 0.12 \times 10^{-9}$ erg~s~cm$^{-2}$. Taking the $F_{\rm bol}$ and $T_{\rm eff}$ together with the {\it Gaia\/} parallax, adjusted by $+0.08$~mas to account for the systematic offset reported by \citet{StassunTorres:2018}, gives the stellar radius as $R = 1.047 \pm 0.031 $~R$_\odot$. 

\begin{figure}
	\includegraphics[width=\columnwidth]{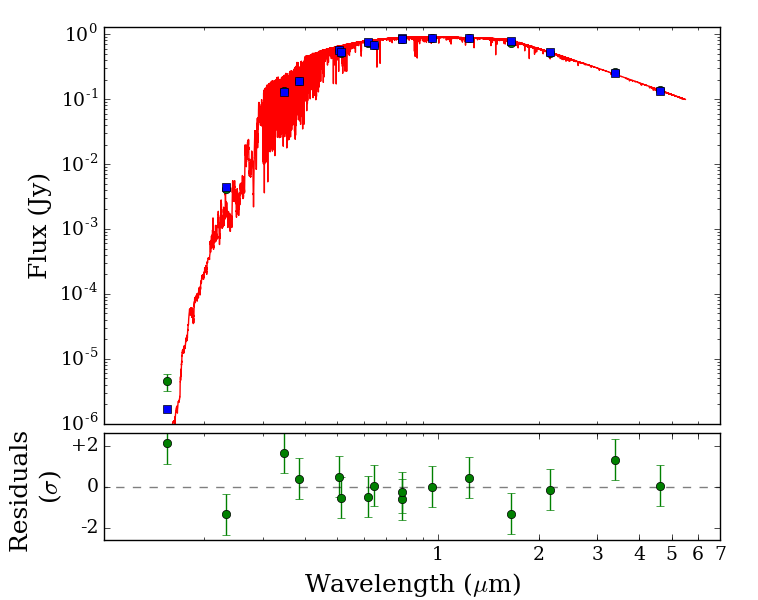}
    \caption{The fitted spectral energy distribution (red line) for \Nplanet\ based on the photometric data (green points) presented in Table~\ref{tab:stellar}. The blue squares show the model flux at the wavelengths of the photometric data.}
    \label{fig:sed}
\end{figure}

We verified the above SED fitting of TOI-222 with the method outlined in \citet{Gillen17}, using the PHOENIX v2 set of models with the information available from blended catalogue photometry and \textit{Gaia} photometric and astrometric data listed in Table \ref{tab:stellar} (see Figure \ref{fig:sed}). Both methods are in very good agreement.

In addition, we can estimate the stellar mass from the eclipsing-binary based empirical relations of \citet{Torres:2010}, which gives $M = 1.07 \pm 0.08 M_\odot$, This, together with the empirical radius above, gives the mean stellar density $\rho = 0.94 \pm 0.10 \rho_\odot = 1.32 \pm 0.14\, \mathrm{gcm}^{-3}$.

\begin{table}
	\centering
	\caption{Stellar Properties of the Primary}
	\begin{tabular}{lcc} 
	Property	&	Value		&Source\\
	\hline
    \multicolumn{3}{l}{Astrometric Properties}\\
    RA		&	\NRA			&2MASS	\\
	Dec			&	\NDec			&2MASS	\\
    2MASS I.D.	& J23563876-4443086	&2MASS	\\
    $\mu_{{\rm R.A.}}$ (\masy) & \NpropRA & UCAC4 \\
	$\mu_{{\rm Dec.}}$ (\masy) & \NpropDec & UCAC4 \\
    \\
    \multicolumn{3}{l}{Photometric Properties}\\
	V (mag)		&\NVmag 	&APASS\\
	B (mag)		&\NBmag		&APASS\\
	g (mag)		&\Ngmag		&APASS\\
	r (mag)		&\Nrmag		&APASS\\
	i (mag)		&\Nimag		&APASS\\
    G (mag)		&\NGAIAmag	&{\em Gaia}\\
    T (mag)     &\NTmag     &\TESS \\
    J (mag)		&\NJmag		&2MASS	\\
   	H (mag)		&\NHmag		&2MASS	\\
	K (mag)		&\NKmag		&2MASS	\\
    W1 (mag)	&\NWmag		&WISE	\\
    W2 (mag)	&\NWWmag	&WISE	\\
    \\
    \multicolumn{3}{l}{Derived Properties}\\

    T$_{\rm eff}$ (K)      & \Nteff               &CORALIE Spectra\\
    $\left[M/H\right]$     & \Nmetal  &CORALIE Spectra\\
    vsini (\kms)	       & \Nvsini			    &CORALIE Spectra\\
    $\gamma_{RV}$ (\kms)   & \Ngamma		        &CORALIE Spectra\\
    log g                  & \Nlogg			&CORALIE Spectra\\
    $R_1$ (\msun)          & $1.047\pm0.031$      &SED fit\\
    $M_1$ (\rsun)          & $1.07\pm0.08$ &TO\\
    $\rho_1$ ($\rho_{\sun}$) & $ 0.94\pm0.17$ &  \\
    Distance (pc)	&  \Ndist	                &BJ \\
	\hline
    \multicolumn{3}{l}{2MASS \citep{2MASS}; UCAC4 \citep{UCAC};}\\
    \multicolumn{3}{l}{APASS \citep{APASS}; WISE \citep{WISE};}\\
    \multicolumn{3}{l}{{\em Gaia} \citep{GAIA}}\\
    \multicolumn{3}{l}{ER = empirical relations using \citet{Benedict16}}\\
    \multicolumn{3}{l}{ and \citet{Mann15}}\\
    \multicolumn{3}{l}{TO = interpolated from \citet{Torres:2010}}\\
    \multicolumn{3}{l}{BJ = from \citet{BJ18}}
    
	\end{tabular}
    \label{tab:stellar}
\end{table}

\subsection{Orbital solution}
\label{sec:solution}
\subsubsection{MCMC framework}
\label{sec:CONAN}
 
We made use of the \emph{\textbf{CO}de for transiting exopla\textbf{N}et \textbf{AN}alysis}, a Markov Chain Monte Carlo (MCMC) framework to carry out a 
simultaneous analysis of transit and radial-velocity data. \emph{CONAN} is based on the transit analysis code 
described in \citet{Lendl17b}, with the added capacities of jointly analysing RV and transit data and thus deriving consistent system properties.
\emph{CONAN} is capable of simultaneously modelling multiple transit and RV data sets, deriving global system properties while allowing for independent baseline models for each data set. 
The MCMC engine can be set to either \emph{emcee} \citep{Foreman-Mackey17} or \emph{MC3} \citep{Cubillos17}. 

To compute transit and RV models, \emph{CONAN} uses the routines of \citep{Kreidberg15} and a Keplerian respectively. The MCMC jump (i.e., fitted) parameters are 
\begin{itemize}
 \item the stellar-to-planetary radius ratio, $R_p / R_\ast$
 \item the timing of mid-transit, $T_0$
 \item the orbital inclination, $inc$
 \item the normalised orbital semi-major axis, $a/R_\ast$
 \item the planetary orbital period, $P$
 \item the RV amplitude, $K$
 \item $\sqrt{e}\sin{\omega}$ and $\sqrt{e}\cos{\omega}$ to derive the eccentricity $e$ and the argument of periastron $\omega$ 
 \item the combination of the systemic RV, $\gamma$, together with the RV zero point $\mathit{RV}_{0,i}$ of each RV data set, $\gamma + \mathit{RV}_{0,i}$
 \item the linear combinations $c_1 = 2u_1 + u2$ and $c_2 = u_1 - u_2$ of the quadratic limb-darkening coefficients $u_1$ and $u_2$ \citep{Holman06}
\end{itemize}

Most photometric time-series are affected by some degree of correlated noise, of observational (e.g. differential effects due to reference stars differing in 
colour), instrumental (e.g. telescope pointing jitter paired with flat field inhomogeneities) or stellar (e.g. rotational variability) origin.
We account for these effects by using parametric baseline models in the form of a combination polynomials or trigonometric functions 
in one or several state variables (e.g., time, stellar FWHM, coordinate shifts, airmass, e.g. see \citealt{Gillon10a,Lendl17b}).
The coefficients can either be found via least-square minimisation at each MCMC step,
or included as MCMC jump parameters. While both approaches fulfil the requirement that baseline coefficients are fit at the same time as the transit parameters (and therefore assure that uncertainties in the baseline parameters are propagated to the inferred transit parameters), the earlier leads to enhanced convergence while being computationally more expensive, the latter allows to trace correlation between the baseline model and transit parameters. Convergence of the MCMC chains is checked with the Gelman \& Rubin test \citep{Gelman92}.

We also include two approaches to compensate for excess noise. Either, the user can choose to scale the errors via the $\beta_{r}, \beta_{w}$ method that consists in comparing the binned and non-binned light curve residuals \citep{Winn08,Gillon10a}, or decide to quadratically add extra noise to each light curve to reach a reduced chi-square equal to unity. Both approaches assure correct weighing between individual data sets. The latter does so on the basis of assuming white Gaussian noise, while the former compensates for excess red noise and often results in $\chi_{\mathit{red}}$ values below unity. Also RV measurements are often affected by excess noise, most prominently due to stellar activity. CONAN allows to account for this \textit{RV jitter} by adding excess noise quadratically to the RV errors such that $\chi_{\mathit{red},RV}^2 = 1$ is reached.

To compute physical properties from the fitted data, CONAN uses input values for $M_\ast$ and $R_\ast$, and draws from Gaussian distributions centred on these
values, and with a user-specified width, for each parameter state output by the MCMC. Here, one may define both $M_\ast$ and $R_\ast$, or define one and fit 
for the other using the information contained in the transit light curve.

\subsection{The TOI-222 system}

We used \emph{CONAN} together with the stellar parameters inferred from the CORALIE data as described in Section \ref{sec:stellar} to derive the properties of the TOI-222 system. Stellar limb-darkening was treated using a quadratic law, with the coefficients derived using the procedures by \citep{Espinoza15}, and held fixed during the analysis. RV data from CORALIE and FEROS were fitted
without the need of including a trend, however the MINERVA RVs were affected by an instrumental drift which we compensated by fitting (individually) quadratic drifts to the RV values of fibres 3,4 and 6 and a linear drift to RVs from fibre 5. The photometric baseline models used were a third-order time polynomial for the NGTS data, and a combination of second order polynomials in the stellar FWHM and the sky flux for the EulerCam data. The previously-detrended \TESS\ data were cut to a shorter sequence of 1-day length centred on the transit and only a constant flux offset was fitted. The light curves are shown in Figure \ref{fig:LCs}. To aid convergence, we first carried out a least-square minimisation and used the resulting values as starting points for the MCMC chains. 

The primary eclipse of TOI-222 is highly grazing, with an impact parameter larger than unity. This inhibits the precise measurement of the secondary star's radius as the radius ratio is strongly correlated with the system's inclination (a larger secondary will create a similarly shallow eclipse when the impact parameter of the eclipse is larger). This means that, albeit the exquisite precision of the photometry, the secondary radius is only loosely constrained.
The parameters of the TOI-222 system are summarised in Table \ref{tab:syspar}.

\begin{table}
\centering                        
\caption{\label{tab:syspar}Properties of the TOI-222 system from a global MCMC analysis.
$^a$ Relative to CORALIE.}
\begin{tabular}{p{4.7cm}p{2.8cm}}       
\hline\hline 
 \multicolumn{2}{l}{Jump parameters} \T  \\
\hline
$T_0$, [BJD - 2450000]                 & $8376.5444_{-0.0006}^{+0.0005} $  \T \\
Radius ratio, $R_1/R_2$                & $0.17_{-0.10}^{+0.37}          $  \T \\
Inclination [deg]                      & $88.1_{-0.84}^{+0.14}          $  \T \\
$a/R_1$                                & $42.7_{-1.8}^{+2.5}            $  \T \\
Period [d]                             & $33.91237_{-0.00013}^{+0.00008}$  \T \\
$\sqrt{e}\sin{\omega}$                 & $0.5018_{-0.0004}^{+0.0001}    $  \T \\
$\sqrt{e}\cos{\omega}$                 & $0.1871_{-0.0002}^{+0.0004}    $  \T \\
RV amplitude $K$ [$\mathrm{kms^{-1}}$] & $13.574\pm0.003                $ \\
\hline
\multicolumn{2}{l}{Primary parameters from spectral analysis}\T\\
\hline
$R_1$ [$R_{\sun}$]             & $1.047 \pm 0.031 $\T\\
$M_1$ [$M_{\sun}$]             & $1.07 \pm 0.08 $\\
\hline
\multicolumn{2}{l}{Derived parameters}\T\\
\hline
$R_2$ [$R_{\sun}$]               & $0.18_{-0.10}^{+0.39}        $\T \\
$M_2$ [$M_{\sun}$]               & $0.23\pm 0.01                $   \\ 
Orbital semi-major axis [au]     & $0.208_{-0.010}^{+0.014}     $   \\
Primary eclipse impact parameter & $1.03_{-0.11}^{+0.37}        $ \T  \\
Orbital eccentricity             & $0.2868_{-0.00027}^{+0.00008}$  \T \\
Argument of periastron [deg]     & $69.55_{-0.05}^{+0.02}       $  \T \\
\hline
\multicolumn{2}{l}{Fixed quadratic limb-darkening parameters}\T \\
\hline
$u_{1,\mathrm{TESS}}$    & $0.3618$\\
$u_{2,\mathrm{TESS}}$    & $0.2180$\\
$u_{1,\mathrm{IC}}$      & $0.3679$\\
$u_{2,\mathrm{IC}}$      & $0.2175$\\
$u_{1,\mathrm{NGTS}}$    & $0.4561$\\
$u_{2,\mathrm{NGTS}}$    & $0.2067$\\
\hline
\multicolumn{2}{l}{Radial velocity zero point offsets$^{a}$, [$\mathrm{ms^{-1}}$] }\T \\
\hline
CORALIE (CERES) & $ -11 \pm 476 $ \\
FEROS & $13 \pm 332 $ \\
MINERVA3 & $5817 \pm 300 $\\
MINERVA4 & $5822 \pm 300 $\\
MINERVA5 & $329 \pm 349 $\\
MINERVA6 & $6153 \pm 293 $\\
\hline
\hline
\end{tabular}
\end{table}


\section{Conclusions and Outlook}
\label{sec:conc}
We present the ephemeris recovery and orbital characterisation of the single-transit TESS candidate TOI-222 by means of radial-velocity monitoring with CORALIE and ground-based photometry with EulerCam and \NGTS. TOI-222 is a binary system with a low-mass eclipsing component orbiting a V=9.3 G2V star in a 33.9\,day orbit, showing $\sim$3500\,ppm-deep grazing eclipses. By observing simultaneously with 10 individual 20\,cm NGTS telescopes, we obtain a photometric precision better than 500\,ppm per 2 minute bin. This demonstrates that we can reach the precision needed to retrieve small-amplitude signals using this technique. While TOI-222 counts as a false positive in the realm of exoplanet discoveries, this detection highlights that long-period single-transit events can be efficiently recovered from the ground, provided well-coordinated high-precision RV and photometric facilities are available. In the light of the several hundreds of predicted single transit candidates from TESS \citep{cooke18, villanueva19}, this opens up a promising avenue for the confirmation of long-period transiting planets.

TOI-222, and future similar discoveries, are particularly exciting as they represent some of the most isolated M-dwarfs with physical properties measured using radial velocities and eclipse photometry. Non-transiting low-mass binaries characterised with radial velocity measurements and/or astrometry require mass-radius-luminosity relations (e.g. \citealt{2000A&A...364..217D} or \citealt{2009MNRAS.394..272S}) which carry relatively large uncertainties and can often be unreliable. Some M-dwarfs around F-/G-type stars have been measured to be hotter and larger than predicted by stellar evolution models. Of those measured within the EBLM project \citep{2012AAS...21934515H}, two stars (WASP$-$30B and J1013$+$01) appear to be inflated, and a third (J0113$+$31) is measured to be $\sim 600\,K$ hotter than expected. A similar result was also seen for KIC1571511 \citep{2012MNRAS.423L...1O} using high-precision optical photometry from the Kepler space telescope. The favoured explanation lies with magnetic activity, as M-dwarfs in binary systems can be kept in fast rotation regime, by tidal-induced spin-orbit synchronisation, generating a enhanced magnetic field via dynamo action \citep{2007A&A...472L..17C}. The effect of this is two-fold: an inhibition of convection in the (almost) fully convective core and higher spot coverage \citep{1966MNRAS.133...85G}. Compared to these objects, TOI-222 has a relatively large orbital period and thus a large orbital separation. 
TOI-222 is also not in synchronous rotation, as the spectroscopically-determined $V\sin\,i$ implies faster rotation, with a period of at least $\sim$14.7~days.
Due to the grazing configuration, the TOI-222 system itself is not suited to probe trends between radius inflation and orbital separation. Similar systems however, which are likely to be found from TESS single transit events in the future, will be instrumental in building a larger sample of well-detatched low-mass eclipsing binaries needed to carry out these studies.

\section*{Acknowledgements}
We thank the Swiss National Science Foundation (SNSF) and the Geneva University for their continuous
support to our planet search programs. 
Contributions at the University of Geneva by FB, LN, ML, OT, and SU were carried out within the framework of the National Centre for Competence in Research "PlanetS" supported by the Swiss National Science Foundation (SNSF).
Based on data collected under the NGTS project at the ESO La Silla Paranal Observatory.  The NGTS facility is operated by the consortium institutes with support from the UK Science and Technology Facilities Council (STFC) under projects ST/M001962/1 and ST/S002642/1. 
The contributions at the University of Warwick by PJW, RGW, DLP, DJA, and TL have been supported by STFC through consolidated grants ST/L000733/1 and ST/P000495/1. DJA acknowledges support from the STFC via an Ernest Rutherford Fellowship (ST/R00384X/1).
The contributions at the University of Leicester by MGW and MRB have been supported by STFC through consolidated grant ST/N000757/1.
CAW acknowledges support from the STFC grant ST/P000312/1.
TL was also supported by STFC studentship 1226157.
MNG acknowledges support from MIT's Kavli Institute as a Torres postdoctoral fellow.
ML acknowledges support from the Austrian Research Promotion Agency (FFG) under project 859724 ``GRAPPA''.
JSJ acknowledges support by Fondecyt grant 1161218 and partial support by CATA-Basal (PB06, CONICYT).
JIV acknowledges support of CONICYT-PFCHA/Doctorado Nacional-21191829, Chile.
RB\ acknowledges support from FONDECYT Post-doctoral Fellowship Project 3180246, and from the Millennium Institute of Astrophysics (MAS).
AJ acknowledges support from FONDECYT project 1171208, and by the Ministry for the Economy, Development, and Tourism's Programa Iniciativa Cient\'{i}fica Milenio through grant IC\,120009, awarded to the Millennium Institute of Astrophysics (MAS).
This project has received funding from the European Research Council (ERC) under the European Union's Horizon 2020 research and innovation programme (grant agreement No 681601).
The research leading to these results has received funding from the European Research Council under the European Union's Seventh Framework Programme (FP/2007-2013) / ERC Grant Agreement n. 320964 (WDTracer).
MINERVA-Australis is supported by Australian Research Council LIEF Grant LE160100001, Discovery Grant DP180100972, Mount Cuba Astronomical Foundation, and institutional partners University of Southern Queensland, UNSW Australia, MIT, Nanjing University, George Mason University, University of Louisville, University of California Riverside, University of Florida, and The University of Texas at Austin.  We respectfully acknowledge the traditional custodians of all lands throughout Australia, and recognise their continued cultural and spiritual connection to the land, waterways, cosmos, and community. We pay our deepest respects to all Elders, ancestors and descendants of the Giabal, Jarowair, and Kambuwal nations, upon whose lands the Minerva-Australis facility at Mt Kent is situated.
Funding for the TESS mission is provided by NASA's Science Mission directorate. We acknowledge the use of public TESS Alert data from pipelines at the TESS Science Office and at the TESS Science Processing Operations Center. Resources supporting this work were provided by the NASA High-End Computing (HEC) Program through the NASA Advanced Supercomputing (NAS) Division at Ames Research Center for the production of the SPOC data products.





\bibliographystyle{mnras}
\bibliography{paper} 








\bsp	
\label{lastpage}
\end{document}